# A COMBINATION BETWEEN VQ AND COVARIANCE MATRICES FOR SPEAKER RECOGNITION[1]

*Marcos Faundez-Zanuy*

Escola Universitària Politècnica de Mataró
Universitat Politècnica de Catalunya (UPC)
Avda. Puig i Cadafalch 101-111, E-08303 MATARO (BARCELONA)
e-mail: faundez@eupmt.es  http://www.eupmt.es/veu

## ABSTRACT

This paper presents a new algorithm for speaker recognition based on the combination between the classical Vector Quantization (VQ) and Covariance Matrix (CM) methods. The combined VQ-CM method improves the identification rates of each method alone, with comparable computational burden. It offers a straightforward procedure to obtain a model similar to GMM with full covariance matrices. Experimental results also show that it is more robust against noise than VQ or CM alone.

## 1. INTRODUCTION

It is well known that attempts to model detail of the characteristics of the speakers' distribution, which may be lost due to changes and distortion in the channel, can lead to nonrobust performance in speaker recognition. A robust approach also attempts to capture detail, but by using multiple simple models rather than a single complicated fragile model [1].

One of the most successful methods for speaker recognition is the Gaussian Model Mixture (GMM) [2], which consists on the modeling of small clusters of speech using one gaussian function for each cluster. Usually nodal, diagonal covariance matrices are used for speaker models. One of the drawbacks of this method [2] is that there is a lower limit on the number of mixtures components necessary to adequately model the speakers. Models must contain at least 16 mixture components to maintain good speaker identification performance. This implies that enough training material is available. It has been found [3] that the error rates with diagonal matrices are always higher than using full covariance matrices, even if we take more distributions (mean vectors). A global covariance matrix also doesn't yield good results.

GMMs, as introduced by Reynolds [2], perform very well but training requires a lot of time and they get numerically unstable when trained with small amount of data. The main problem is the inversion of the (underestimated) covariance matrices. Pure vector quantization, e.g. using k-means clustering or the LBG algorithm by Linde, Buzo and Gray [4], on the other hand is numerically stable and rather fast, but the performance in speaker recognition is not as good as for GMMs.

While the GMM algorithm implies an iterative procedure that simultaneously optimizes via Expectation-Maximization the mean and variance of the gaussians, our procedure is straight forward an lets a more sophisticated committee of experts between several classifiers.

On the other hand, a recent paper [5] has shown that the GMM without the EM iteration can outperform the GMM with EM when there is a mismatch between training and testing conditions. This is because a more robust model can be obtained if overtraining is avoided. It presents an about 16% improvement over the baseline performance of the EM trained diagonal GMM with a comparable number of parameters, with an easier algorithm.

In this paper, we present a new algorithm for speaker recognition that combines the mean and variances of several clusters, estimated with the classical VQ and CM methods.

This paper is organized in the following way: section 2 summarizes the VQ and CM methods and presents our new method. Section 3 presents the experimental results, with special emphasis on the comparison between the classical method and the new one. It also introduces the conditions of the experiments and the used database. Finally, section 4 is devoted to the main conclusions.

---

[1] This work has been supported by the CICYT TIC97-1001-C02-02

## 2. SPEAKER RECOGNITION USING VQ AND CM

This section summarizes the VQ and CM algorithms, presents the new VQ-CM algorithms, and the conditions of the experimental results.

### 2.1 Speaker recognition using VQ

In this system, each speaker is modeled with a vector quantizer during the training process. The identification is done quantizing the input sentence with all the codebooks and choosing the quantizer that gives the lowest accumulated error. A detailed explanation of this system can be found in [6].

The number of parameters used in each model is:

$$parameters = 2^{N_0} \times P$$

Where P is the analysis order of the parameterization (dimension of LPCC vectors in our study) and $No$ is the number of bits of the codebook.

### 2.2 Speaker recognition using CM

A covariance matrix (CM) is computed for each speaker, and an Arithmetic-harmonic sphericity measure is used in order to compare matrices [7]:

$$\mu(C_j C_{test}) = \log\left(tr(C_{test}C_j^{-1})tr(C_j C_{test}^{-1})\right) - 2\log(P)$$

Where $tr$ is the trace of the matrix.

The number of parameters for each speaker is $\frac{P^2 + P}{2}$ (the covariance matrix is symmetric).

For the VQ models, more parameters imply a bigger codebook, while for the CM implies a higher dimensional cepstral vectors.

### 2.3 Proposed VQ-CM method

This new method consists on the following steps:

Enrollment phase

1. To compute a codebook of 1 or 2 bits.
2. To cluster the training vectors around each centroid.
3. To compute one covariance matrix for each cluster.

Test phase

1. To cluster the test sequence vectors, and to compute $d_0$.
2. To compute one covariance matrix for each cluster of the test sequence, and to compute $d_i \quad i=1,\ldots N$

Where we have used the following notation:

$d_0$= distance from the test vectors to the VQ classifier.

$d_i$= distance from the CM matrix of test vectors of cluster $i$ to the CM of the training vectors of cluster $i$, $i=1,\ldots N$.

This method implies the use of $N+1$ classifiers (each covariance matrix of each cluster plus the VQ classifier) that can be combined in several ways in order to improve the results of each classifier alone.

The number of parameters of the VQ-CM method is:

$$parameters = 2^{N_o} \times \left(P_1 + \frac{P_2^2 + P_2}{2}\right)$$

Where $P_1$ is the dimension of the vectors of the VQ, while $P_2$ is the dimension of the vectors of the CM.

It is interesting to observe that the estimation of the covariance matrices is based on the data vectors associated to his center only. This leads to a segmentation of the feature space. In the GMM case, all data vectors are used for the estimation of every mean vector and covariance matrix. No segmentation of the feature space is done. Thus, the computation of each covariance matrix is faster than the GMM or CM computation, because the number of associated vectors is smaller (the training sequence has been clustered using the vector quantizer).

Figure 1 shows a simplified scheme of the new proposed method.

### 2.4 Database

Our experiments have been computed over 43 speakers from the Gaudi database [8] that has been obtained with a PC connected to an ISDN. Thus, the speech signal is A law encoded at 8kHz an 8 bit/sample. The speech signals are pre-emphasized by a first order filter whose transfer function is H(z)=1-0.95z$^{-1}$. A 30 ms Hamming window is used, and the overlapping between adjacent frames is 2/3. A cepstral vector of order 16 ( $P_1 = 16$ ) was computed from the LPC coefficients. One minute of read text is used for training, and 5 sentences for testing (each sentence is about two seconds long).

M. Faundez-Zanuy, "A combination between VQ and covariance matrices for speaker recognition," *2001 IEEE International Conference on Acoustics, Speech, and Signal Processing. Proceedings (Cat. No.01CH37221)*, 2001, pp. 453-456 vol.1, doi: 10.1109/ICASSP.2001.940865.
## 3. RESULTS

Table 1 compares the obtained results with the classical VQ and CM methods alone against the new proposed method (VQ-CM) for 2 and 4 clusters, and covariance matrices of size 10×10 in each cluster. It shows the number of parameters in each model, and the identification rates. The computational burden is also included during training and testing phases.

| method | flops training [M] | flops test [M] | Param. | Ident (%) |
|---|---|---|---|---|
| VQ (No=1) | 2.07 | 0.72 | 32 | 86.5 |
| VQ (No=2) | 4.5 | 1.42 | 64 | 92.1 |
| VQ (No=6) | 58 | 22.7 | 1024 | 97.7 |
| CM (10×10) | 0.73 | 0.04 | 55 | 89.3 |
| CM (20×20) | 2.89 | 0.16 | 210 | 95.8 |
| VQ-CM (No=1) | 2.07+0.9 | 0.08 | 142 | 97.9 |
| VQ-CM (No=2) | 4.5+2.4 | 0.11 | 284 | 99.1 |

Table 1: Comparison between the classical VQ and CM classifiers, and the proposed VQ-CM method.

Table 2 compares several classifiers. It is interesting to observe that the covariance matrix of each cluster is a good classifier that yields similar results to the classical VQ and CM methods. Obviously, the committee of several classifiers improves the results of each classifier alone (see chapter 7 of [9]).

| Classifier | No=2 | No=1 |
|---|---|---|
| $d_0$ | 92.1 | 86.5 |
| $d_1$ | 97.2 | 94.4 |
| $d_2$ | 95.4 | 96.3 |
| $d_3$ | 94.4 | |
| $d_4$ | 90.7 | |
| $\sum_{i=0}^{N} d_i$ | 99.1 | 98.6 |
| $\sum_{i=1}^{N} d_i$ | 99.1 | 97.7 |
| $median(d_i)\big|_{i=0}^{N}$ | 99.1 | 96.3 |

Table 2: Identification rates for several combination schemes.

Table 3 compares the identification rates of the proposed scheme against the classical VQ and CM alone, for several SNR ratios. It can be seen that the combination of several small covariance matrices (VQ-CM algorithm) is more robust against noise, for a small number of clusters ($N_o = 1$).

| SNR→ | ∞ | 30 dB | 25 dB | 20 dB | 15 dB |
|---|---|---|---|---|---|
| VQ No=6 | 97.7 | 97.67 | 95.81 | 81.86 | 48.84 |
| CM 20×20 | 95.8 | 95.81 | 94.88 | 91.16 | 69.3 |
| VQ-CM No=1 | 97.7 | 97.67 | 95.35 | 90.23 | 73.49 |
| VQ-CM No=2 | 99.1 | 95.81 | 85.12 | 57.67 | 28.84 |

Table 3: Identification rates for several SNR.

A voting scheme has also been tested, but the results are worst that the combination of distance measures.

Table 4 shows the results for several SNR ratios and different sizes for the covariance matrices, for two clusters (No=1). The global number of parameters is also included.

| $P_2$ → | 2 | 4 | 6 | 8 | 10 | 12 |
|---|---|---|---|---|---|---|
| SNR=∞ dB | 29.8 | 93.5 | 94 | 97.7 | 97.7 | 99.1 |
| SNR=30 dB | 28.83 | 91.63 | 93.95 | 96.74 | 96.28 | 98.14 |
| SNR=25 dB | 26.04 | 89.77 | 91.16 | 95.81 | 95.81 | 96.28 |
| SNR=20 dB | 21.86 | 86.05 | 82.79 | 88.84 | 91.16 | 90.7 |
| SNR=15 dB | 14.88 | 67.91 | 68.84 | 75.81 | 74.42 | 72.56 |
| Parameters | 38 | 52 | 74 | 104 | 142 | 188 |

Table 4: Identification rates for several SNR and P values, for No=1.

## 4. CONCLUSIONS

In this paper, we have presented a new algorithm for speaker recognition that combines VQ with covariance matrices. Main conclusions are:

- VQ-CM offers a good compromise between computational complexity, number of parameters in each model, and identification results. It outperforms the classical VQ and CM algorithms with a comparable complexity burden. In our simulation results, the identification rates are 1.4 and 3.3% higher than the VQ and CM results respectively.

- VQ-CM is a combination of small models that achieves more robust performance against noise, especially for

low SNR. In these cases, the drop in the recognition rates is smaller than for the classical VQ and CM methods (identification rate equal to 73.5% for the VQ-CM method versus 69.3 and 48.8 of CM and VQ respectively, when the SNR is 15dB)

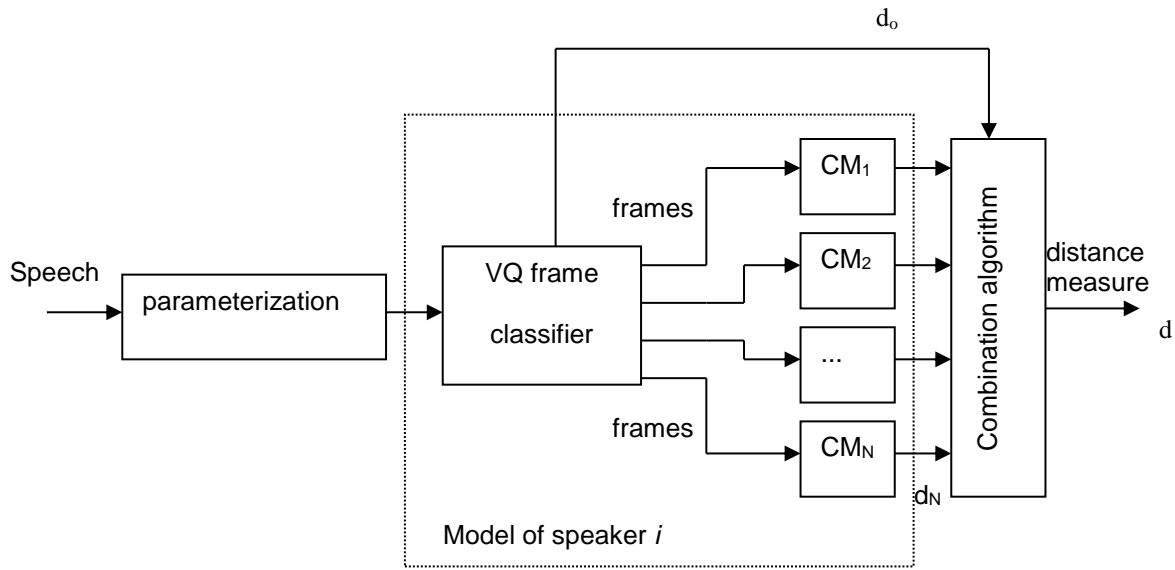

Figure 1: Scheme of the proposed model for each speaker.

___

**ICASSP'01**
**Paper title: A COMBINATION BETWEEN VQ AND COVARIANCE MATRICES FOR SPEAKER RECOGNITION**

**Categories:**
3.11   Speaker identification          8.4

**What problem have you addressed? Why is the problem important?**
Speaker recognition is one of the most promising biometric identification systems, because it has the advantage that can be easily used for identification through the telephone network. In this paper we propose the combination of Vector Quantization and Covariance Matrix methods in order to improve the results of each method alone with a similar computational burden. The basis of this proposition is that a combination of small models gives better results than a complicated model with a lot of parameters.

**What is the original contribution of this work?**
The original contribution of this work is the presentation of a new method for speaker recognition that uses the means and variances of the feature vectors with reduced computational burden. Simulation results show that this method is more robust against noise (the identification rate of the VQ-CM is 73.5% versus 69.3 and 48.8 of CM and VQ respectively, when SNR=15 dB).